\pdfoutput=1
\documentclass[12pt]{iopart}

\usepackage{graphicx}
\usepackage{color}
\usepackage[english]{babel}

\usepackage{setspace}

\begin{document}
 \title[Using smartphone pressure sensors to measure vertical velocities...]
{Using  smartphone pressure sensors to measure vertical velocities of elevators, stairways, and drones}
 
\author{Mart{\'i}n Monteiro$^1$, Arturo   C. Mart{\'i}$^2$ }

\address{$^1$ Universidad ORT Uruguay}

\address{$^2$ Facultad de Ciencias,
  Universidad de la Rep\'{u}blica, 
  Uruguay}

	     
\ead{marti@fisica.edu.uy}

\begin{abstract}
We measure the vertical velocities of elevators, pedestrians climbing stairs, and drones (flying unmanned aerial vehicles), by
means of smartphone pressure sensors. The barometric pressure obtained with the smartphone is related  
to the altitude of the device via the hydrostatic approximation. From the altitude values, vertical velocities are derived. The approximation considered is valid in the first hundred
meters of the inner layers of the atmosphere. In addition to pressure, acceleration values were also recorded using the built-in accelerometer.
Numerical integration was performed, obtaining both vertical velocity and altitude.
We show that data obtained using the pressure sensor is significantly less noisy than that obtained using the accelerometer.
Error accumulation is also evident in the numerical integration of the acceleration values. In the proposed experiments, 
the pressure sensor also outperforms GPS, because this  sensor does not receive satellite signals indoors and, in general, the operating frequency is considerably lower than that
of the pressure sensor.  In the cases in which it is possible,  comparison  with reference values taken from the architectural plans of buildings validates 
the results obtained using the pressure sensor. 
This proposal is ideally performed as an external or outreach activity with students to gain insight about fundamental questions in mechanics, fluids,
and thermodynamics.
\end{abstract}

\noindent{\it Keywords\/}:  smartphone, pressure sensor, drone, accelerometer, barometer

\pacs{01.50.Pa, 
01.40.My,  
92.60.-e} 


\maketitle
\section{Introduction}

Smartphone use in physics courses has expanded considerably in recent years. 
Because of the availability of several sensors in these modern devices 
(accelerometers, gyroscopes, magnetometer, etc.),  several experiments can be implemented in laboratories (see past volumes
from this journal or the column iPhysicsLab 
in Physics Teacher). The proposed  experiments span a wide range of topics including mechanics 
\cite{Chevrier2013,monteiro2014angular,monteiro2014exploring,MONTEIRO2015,countryman2014familiarizing,vieyrafive,gonzalez2015doing},
oscillations \cite{castro2013using}, waves \cite{Kuhn2013acoustic,monteiro2015measuring} , electricity \cite{Forinash2012} , magnetism \cite{0143-0807-36-6-065002,lara2015demonstrations},
and modern physics \cite{kuhn2014iradioactivity}.
The most important advantages of using  smartphones are both increasing availability among young people and the decreasing cost
(in comparison  with other sensors which are specifically designed for teaching purposes).
It is also worth mentioning the possibility of making simultaneous measurements with several sensors such as gyroscopes and accelerometers 
\cite{monteiro2014angular,monteiro2014exploring,MONTEIRO2015}.
One additional advantage is that experiments with smartphones can be easily performed in non-traditional locations such as playgrounds \cite{monteiro2014angular},
amusement parks, mechanical \cite{moll2010amusement,bagge2002classical,vieyra2014analyzing} or water parks \cite{cabeza2014learning}, or travel facilities \cite{kinser2015relating}. 
Every year, new smartphone models with new capabilities are released; therefore, we expect more physics experiments will be proposed for them.

One capability of smartphones which has received little attention is the use of their pressure sensor, or barometer, 
\cite{monteiro2016exploring,vanini2016using}.  This sensor has been recently included in smartphones. In particular, 
the popular Samsung Galaxy S III, launched in 2012, was one of the first to include a barometer. Concerning the Apple brand, such sensors are included on the iPhone 6 series, 
the Air 2, and Pro and Mini 4 iPad models.  In a very recent  experiment \cite{monteiro2016exploring}, the characteristics of the inner layer  of the atmosphere 
(the  first hundred meters)   were analyzed  using a smartphone mounted on an unmanned aerial vehicle (UAV) or \textit{drone}.
In this experiment, the pressure was obtained using the smartphone's pressure sensor and compared to different models of the 
atmosphere.
In a recent study, a methodology based on the sensors commonly available on smartphones and tablets was proposed 
for the automatic recognition of major vertical displacements in human activities  \cite{vanini2016using}. 
One additional related study included the use of a pressure sensor mounted on an Arduino  
board (an open-source project that created microcontroller-based kits for building digital devices and interactive objects)
 \cite{carvalho2014observing,silva2015open} to study barometric variations in atmospheric variables.

An  exhaustive study of the properties of  pressure sensors included in modern devices and the characteristics of 
measurements taken inside buildings is presented in  \cite{muralidharan2014barometric}.
In this study, results show that  pressure readings show significant time-of-day variations; the 
difference in pressure across different floors of a building 
is remarkably consistent and steady for any given building and is highly robust to changes in the phone's on-body placement 
and orientation, making it a significantly more robust sensor  for real-world vertical activity detection than the accelerometer.

In this paper, we focus on the use of pressure sensors, or barometers, which are built into  contemporary smartphones to 
obtain vertical velocity data in several contexts:  in the elevators of tall buildings, in pedestrian climbing stairways,  and on \textit{drones}.
From the obtained pressure values, using the hydrostatic approximation,  the altitude and the ascending or descending speed is obtained.
Our results are compared with the reference values provided by architectural plans or maximal
ascending  or descending speed reported by the manufacturer of the \textit{drones} (and are in very good agreement).
In previous approaches, using smartphones placed on the floor of an elevator,  vertical acceleration is registered  and subsequently integrated to
obtain the time-dependent height  \cite{kinser2015relating}. However, the intrinsic noise of the acceleration sensor or the external 
vibration induced by the elevator cable  \cite{kuhn2014analyzing} or the propellers makes it difficult to obtain accurate vertical speed data
\cite{monteiro2015analyzing}.

The proposed activity provides a good opportunity to gain insight into the physical characteristics of the inner atmosphere and to review the thermodynamics
properties involved, such as pressure and temperature.
One notable advantage of the present proposal is that it can be performed by students as an external activity. 
The only material needed is a smartphone with a pressure sensor, also called a barometer.

This paper is organized as follows. In the next section, we define some of the properties of 
the atmosphere and how pressure is related to altitude.
The experiments are described in Sec.~\ref{sec:exp}.
Finally, in Sec.~\ref{sec:con} some discussion and several concluding remarks are presented.

\section{Inner layers of the atmosphere}
\label{sec:theo}

The atmosphere is an extremely complex system; usually, to simplify its study, it is divided into several layers. The inner layer,
the first $11$ km of height, is the  \textit{troposphere}, in which the  temperature has a roughly linear gradient, called the lapse rate,
whose standard value is $ –0.0065 \,^o$C/m.

In a previous experiment \cite{monteiro2016exploring}, using a smartphone mounted on a drone,
we obtained pressure  from a barometer and altitude from a GPS device and, using a linear fit,
compared the value of the  density to the  standard values available in the literature. 
The experiment showed that, in the inner layer of the atmosphere, constant 
density is a good approximation. In this case, the hydrostatic approximation can be employed
\begin{equation}
P(z)=P_0 -  \rho  g z 
\end{equation}
where $P(z)$ is the pressure at height $z$, $P_0$ is the pressure at  $z=0$, $\rho$ is the air density, and  $g=9.8$ m/s$^2$ is the gravitational acceleration. 
The density can be obtained by applying the ideal gas law
\begin{equation}
\rho = \frac{PM}{TR}
\end{equation}
where $M=0.029$ kg/mol is the molar mass of the air (as a mixture of gases), $T$ is the absolute temperature, and $R=8.31$ J/mol$\cdot$K  is the gas constant. 
Then, the altitude $z$ can be easily obtained as
\begin{equation}
z=(P_0 - P) \frac{T R}{ P M g}
\end{equation}

In the first layers of the atmosphere, the temperature decreases approximately $0.2\%$ per 100-m increase in altitude, and the pressure decreases 1\% over the same change. 
Then, using the previous equation, we estimate that the density decreases by less than 0.8\% per 100 m. Under these assumptions, 
up to first order (i.e., neglecting higher order terms), we  obtain a linear relationship between pressure and altitude
\begin{equation}
z=(P_0 - P) \frac{T_0 R}{ P_0 Mg }
\label{eq:zP}
\end{equation}
where $T_0$ is the ambient temperature at $z=0$. An expression to obtain the vertical velocity can easily be derived.
\section{Experiments}
\label{sec:exp}

\begin{figure}
\begin{center}
\centerline{\includegraphics[width=0.26\textwidth]{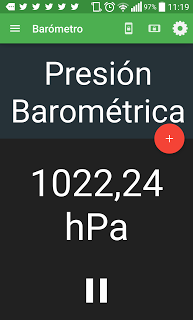}}
\caption{Screenshot of the Physics Toolbox \textit{app} showing the barometric pressure \textit{(Presi\'on barom\'etrica)}.}
\label{fig:01}
\end{center}
\end{figure}

We measured the barometric pressure in different situations: moving elevators, flying \textit{drones}, and pedestrians climbing stairways.
The experiments were performed using the pressure sensors of two different smartphones: an LG model G3 and a Samsung Galaxy S6. 
The \textit{apps} used to register both the pressure and the vertical acceleration, 
were Physics Toolbox Suite (www.vieyrasoftware.net) and AndroSensor  (www.fivasim.com/androsensor.html).
A screenshot of the Physics Toolbox Suite is shown in Fig.~\ref{fig:01}.  After performing the experiments, data is exported to a
spreadsheet and analyzed.

\subsection{Vertical velocities of elevators}

\begin{figure}
\begin{center}
\includegraphics[width=0.562\textwidth]{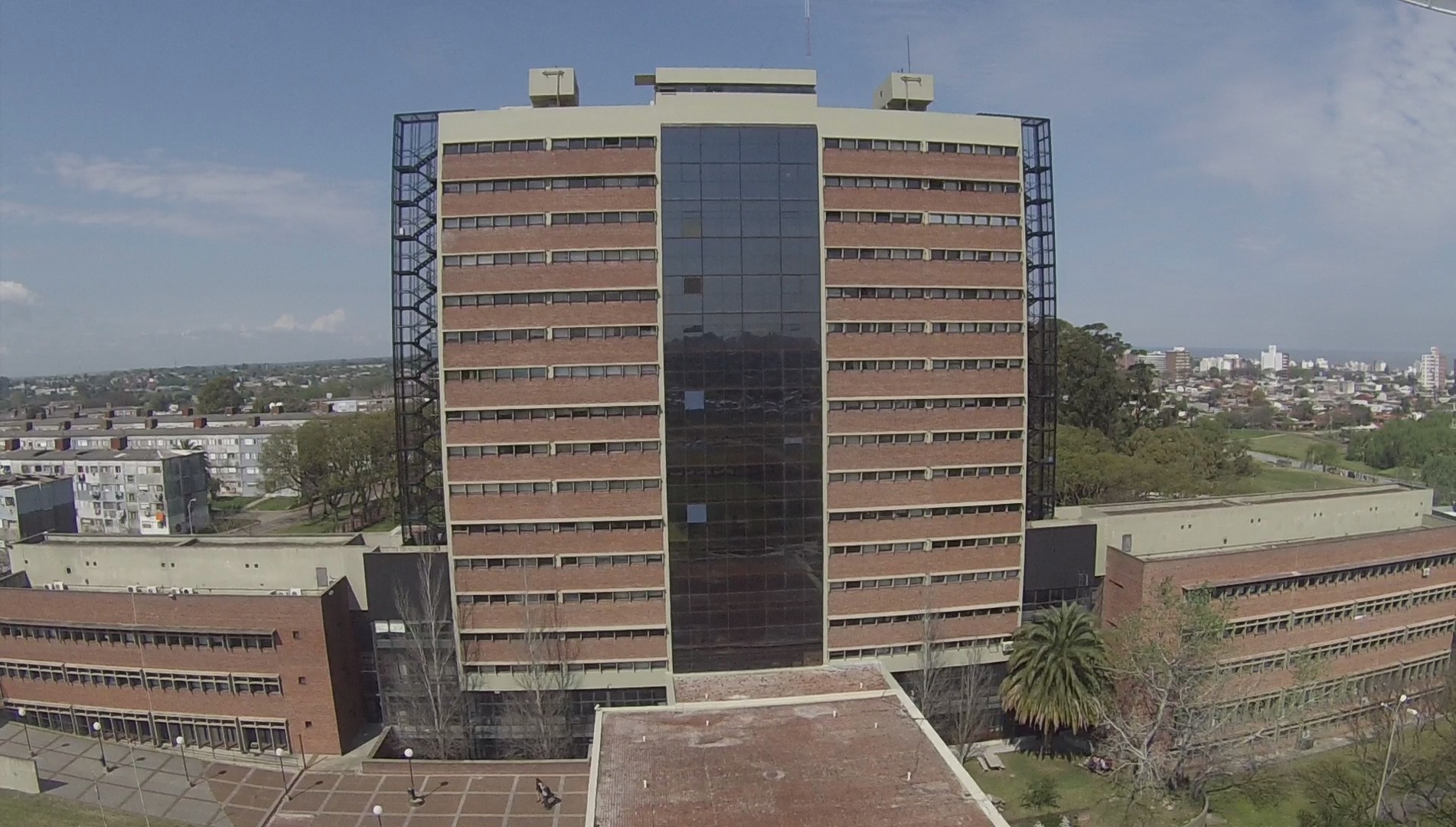}
\includegraphics[width=0.425\textwidth]{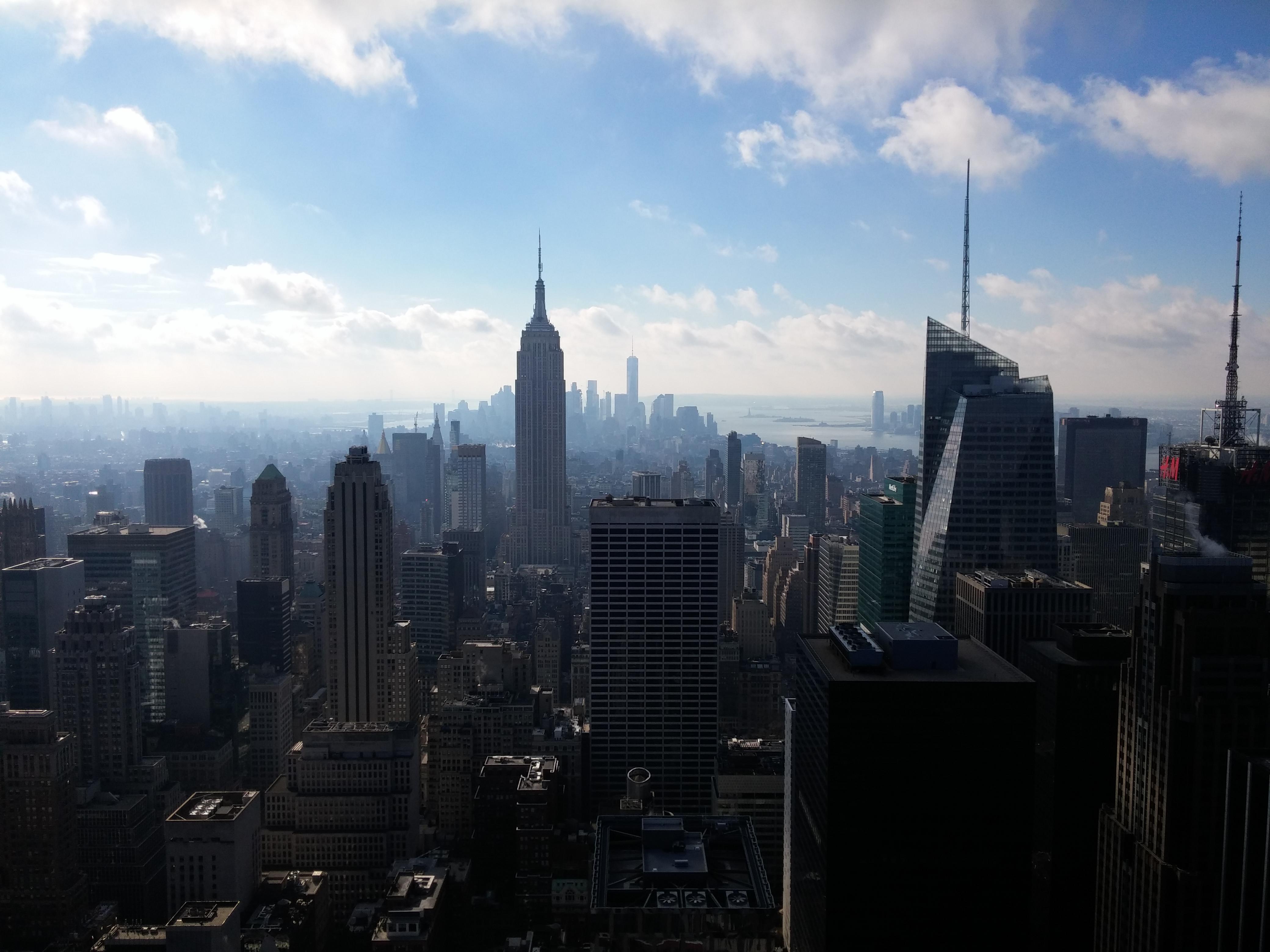}
\caption{  An aerial photograph, taken from the drone, of the main building of the School of Sciences (Universidad de la Rep\'ublica) located in Montevideo, Uruguay (left), and 
a typical  view of New York City, USA, taken from the Top of the Rock observation deck, including the iconic Empire State Building
at the center of the image (right). Photographs taken by the authors.}
\label{fig:fcien-vista}
\end{center}
\end{figure}

Vertical velocities were obtained in three buildings, one in Montevideo (Uruguay) and two in New York City (USA). 
The first building is the Sciences School (FCIEN) in Uruguay, see Fig.~\ref{fig:fcien-vista},
and the other two are 
skyscrapers in New York City: the \textit{Empire State Building} (ESB) and 30 Rock (30R) at Rockefeller Center. 

\begin{figure}
\begin{center}
\includegraphics[width=0.950\textwidth]{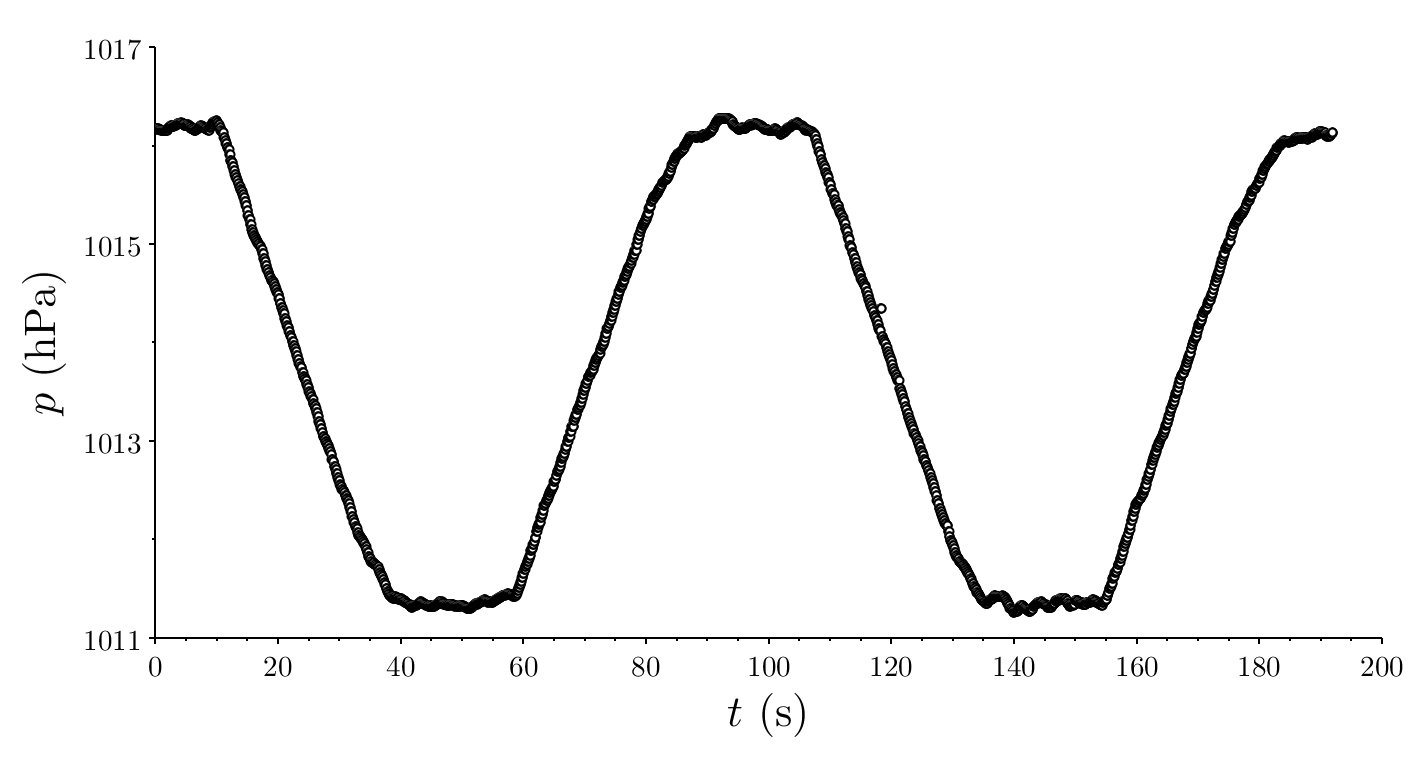}
\caption{Pressure as a function of time during two round trips between the ground floor and the 15th floor in  FCIEN's  elevator.
The ambient temperature was $(T_0=293 \pm 1)$ K.
}
\label{fig:fcien-pressure17}
\end{center}
\end{figure}

\begin{figure}
\begin{center}
\includegraphics[width=0.850\textwidth]{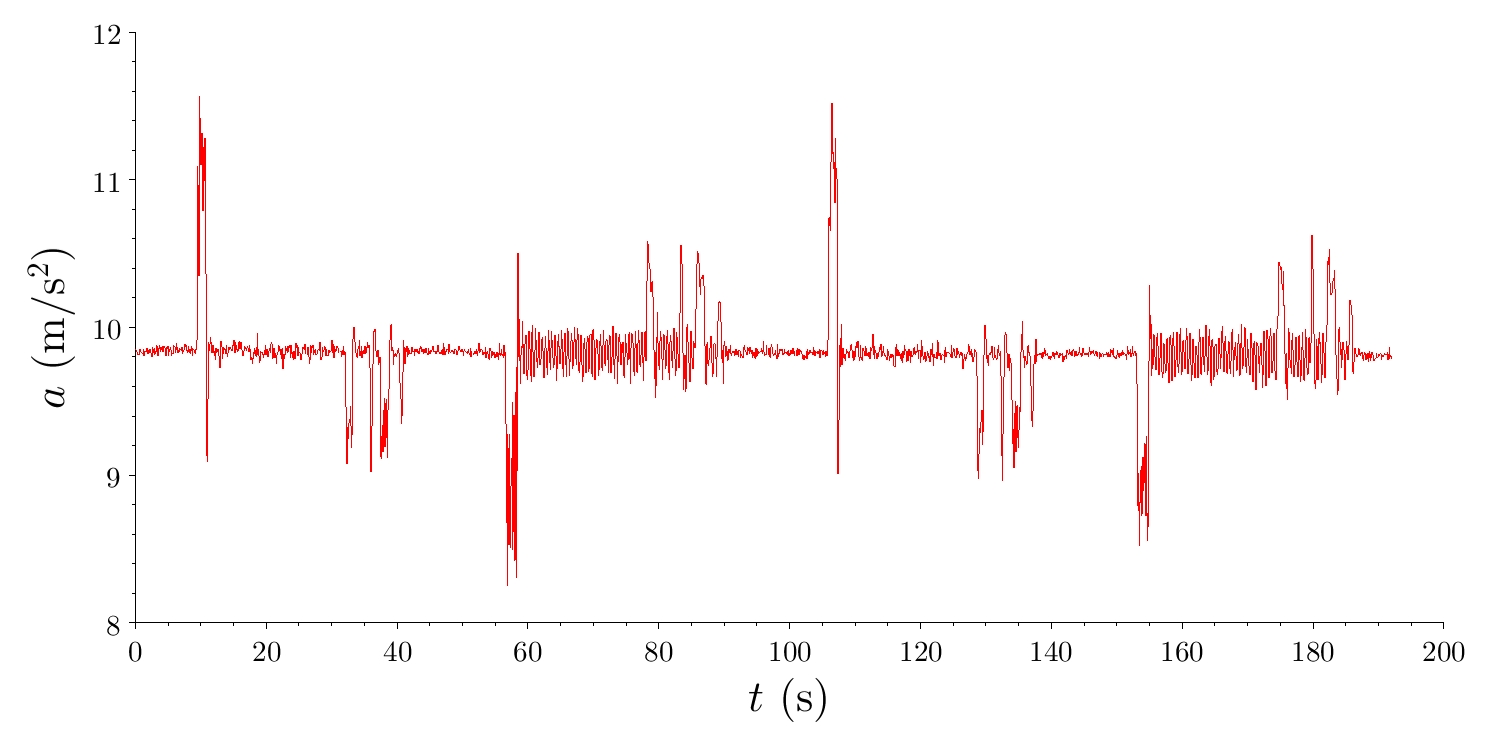}
\includegraphics[width=0.850\textwidth]{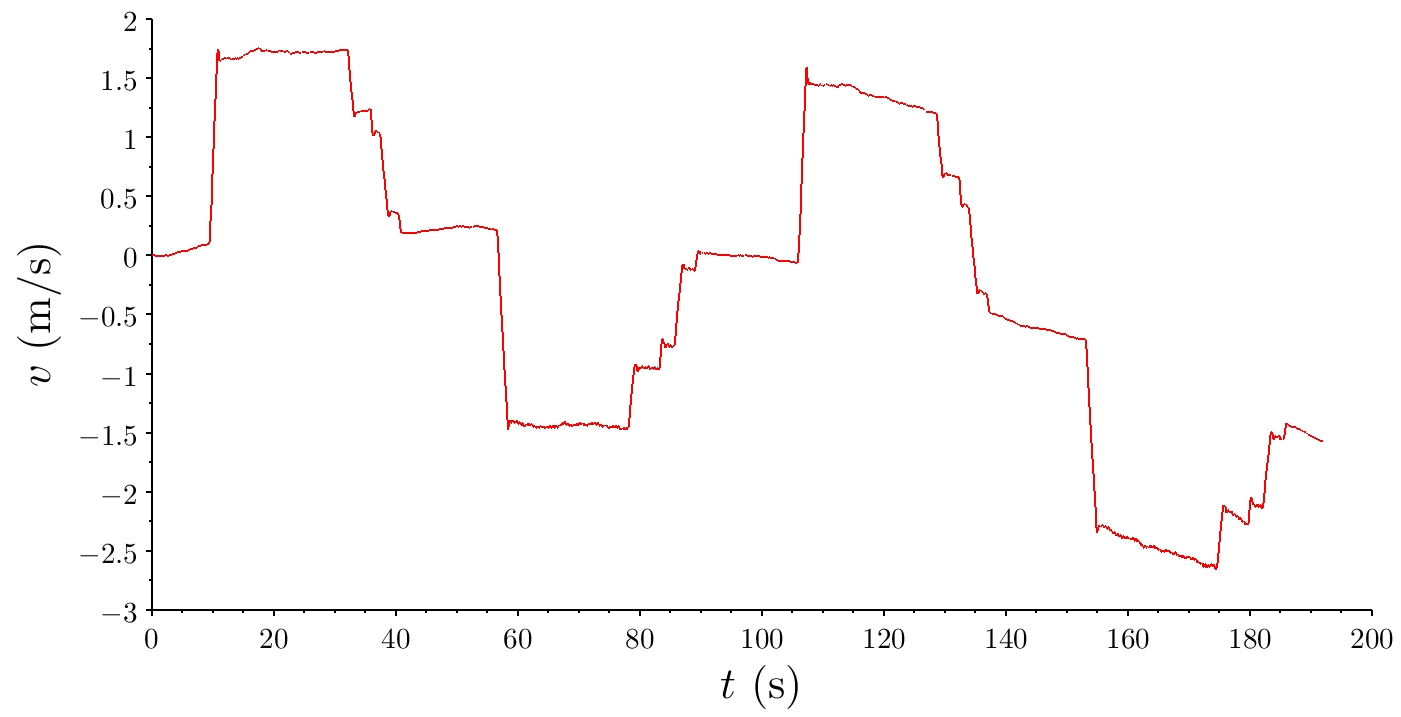}
\caption{Vertical acceleration (top) and velocity (bottom) recorded in the FCIEN elevator. 
The velocity was obtained by numerically integrating the acceleration values. Note that the velocity does not return to zero even when the 
elevator is at rest.}
\label{fig:fcien-acceleration17}
\end{center}
\end{figure}

During two round trips between the Physics Institute (ground floor) and the 15th floor in FCIEN, the barometric pressure and 
the vertical acceleration were recorded at a sampling rate  of $20$ Hz. The Samsung Galaxy smartphone was lying on the floor 
of the elevator throughout the experiment. The ambient temperature was measured using a thermocouple. Small temperature variations, i.e.,
$\pm 1$ K, were detected based on the current floor and the height with respect to the elevator's floor. The average temperature value was used.
 The pressure values are given in Fig.~\ref{fig:fcien-pressure17}.
The accelerometer values are given in Fig.~\ref{fig:fcien-acceleration17} (top); the vertical velocity, numerically integrated
using an Euler scheme, is displayed in the bottom panel.

\begin{figure}
\begin{center}
\includegraphics[width=0.950\textwidth]{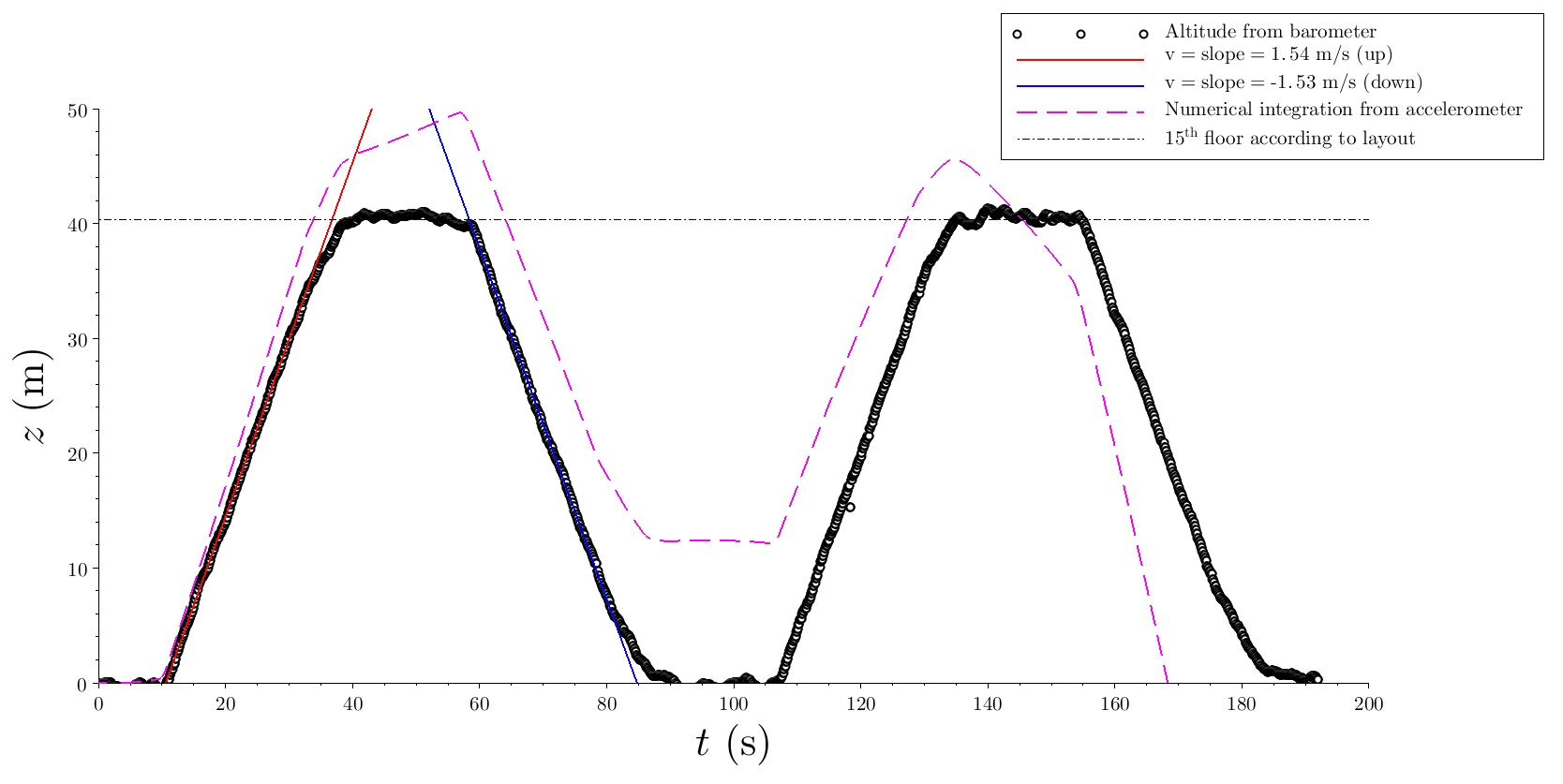}
\caption{Altitude values at the elevator of FCIEN obtained from measurements by the pressure sensor (symbols) and twice integrating the 
numerical values of the acceleration. The true height of the 15th floor, obtained from the architectural plans, is indicated by a dot-dashed line. 
The vertical velocities of the elevator during one ascending and one descending trip obtained using a linear fit are indicated. Note that the velocities
 are similar for all trips.
}
\label{fig:fcien-altitude17}
\end{center}
\end{figure}

The elevator's altitude obtained from the pressure measurements and Eq.~\ref{eq:zP} is shown in Fig.~\ref{fig:fcien-altitude17}. 
The actual height of the 15th floor, verified using the architectural plans, is indicated by a dot-dashed horizontal line.
To compare the use of the pressure sensor with that of the acceleration sensor, the velocity, 
shown in Fig.~\ref{fig:fcien-acceleration17} (bottom) and the altitude, indicated by a green line in Fig.~\ref{fig:fcien-altitude17}, were also obtained. 
Two facts are particularly important. First, there was very good agreement between the measurements obtained using the pressure sensor values and the reference values.
Second, there was a significant discrepancy when altitude values were obtained by integrating the acceleration.
We claim two factors as possible origins of this discrepancy. One source of uncertainty is the intrinsic noise of the accelerometer.
The other source is the accumulation of errors during numerical integrations. A small uncertainty in the acceleration at the beginning of the 
experiment is propagated and amplified throughout the experiment. In addition, because the accelerometers are actually \textit{force} sensors \cite{MONTEIRO2015}, they are 
centered at gravitational acceleration, instead of a zero value; therefore, small uncertainty in thegravitational acceleration causes significant uncertainty in the calculation of the altitude at later times.
This problem is intrinsic to the method and cannot be solved by changing the sampling rate. 
In the present experiment, several sampling rates were evaluated and the best value was chosen, as shown below.

\begin{figure}
\begin{center}
\includegraphics[width=0.490\textwidth]{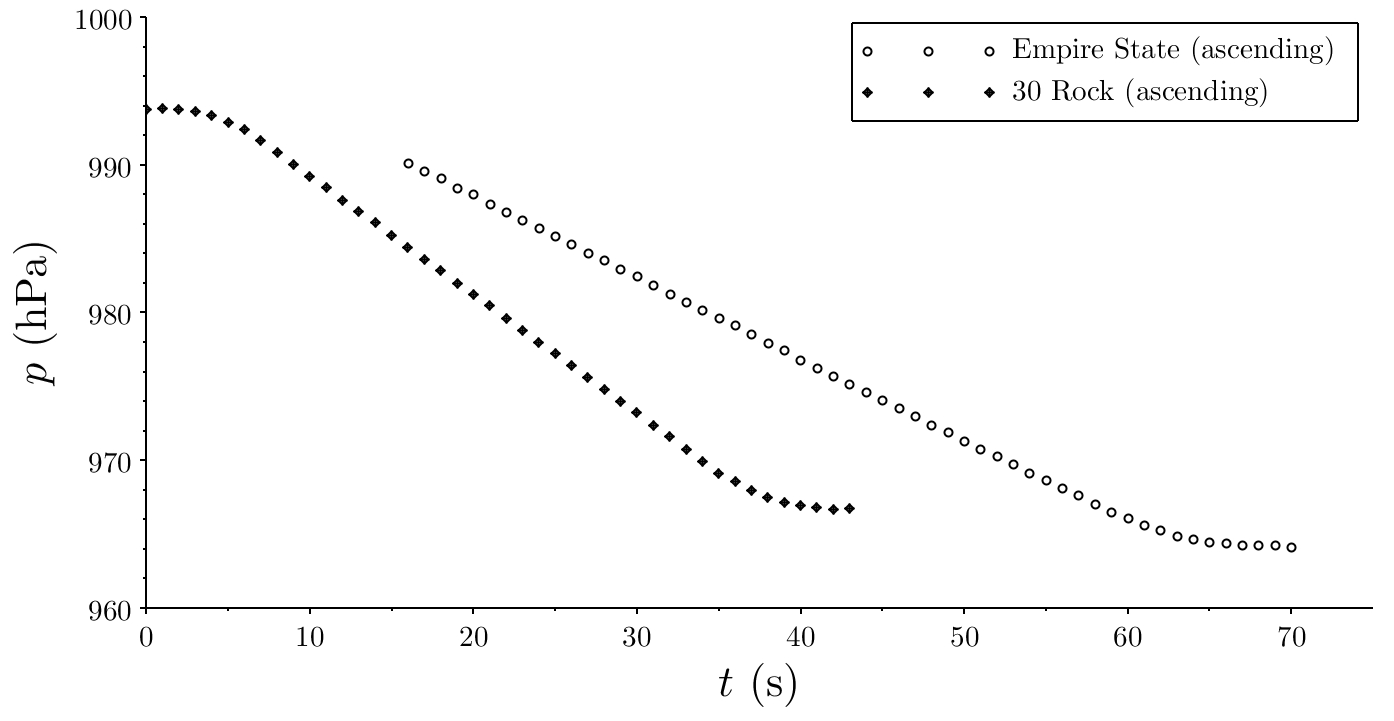}
\includegraphics[width=0.490\textwidth]{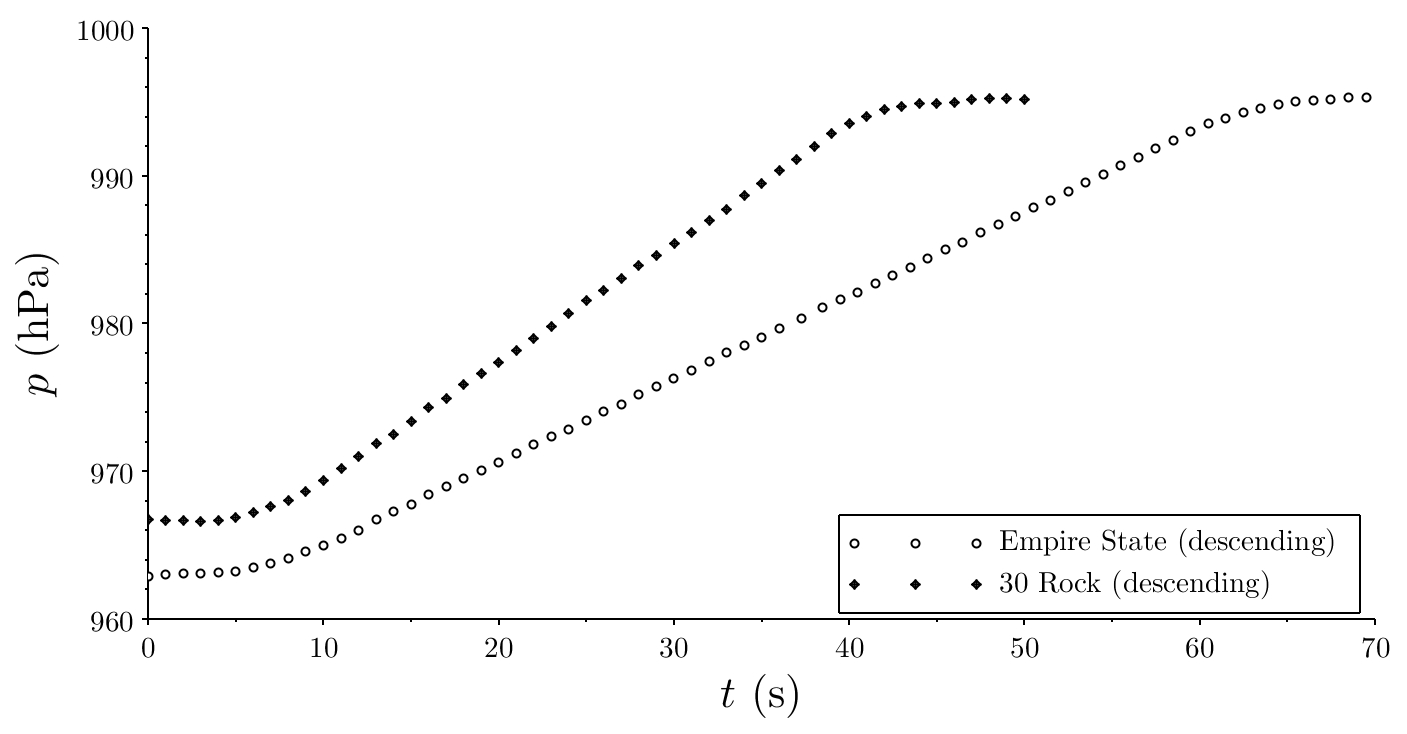}
\caption{Pressure in the elevators in the ESB and in 30R during ascending (left) and 
descending (right) trips.}
\label{fig:pre}
\end{center}
\end{figure}

In Fig.~\ref{fig:pre}, the temporal evolution of the pressure for ascent and descent in both the ESB
and 30R is depicted.   In these experiments, the LG smartphone was held horizontally in the hands of one of the authors.
As in previous figures, the decrease (increase) of the pressure is clearly seen while the elevator is 
ascending (descending).  The ascent intervals last approximately $40$ s in 30R and $60$ s in the ESB.
The elevator speed can be obtained from the pressure using the relationship between atmospheric pressure and altitude, given in Eq.~\ref{eq:zP}.
In addition, the vertical velocities of the elevators are calculated based on the slope of the previous graph.  
From these measurements, we note that speeds of ascent and descent are very similar in both elevators. 
However, the speed of the elevator in 30R (approximately $25$ km/h) is greater than that in the ESB (approximately $17$ km/h).

\begin{figure}
\begin{center}
\includegraphics[width=0.490\textwidth]{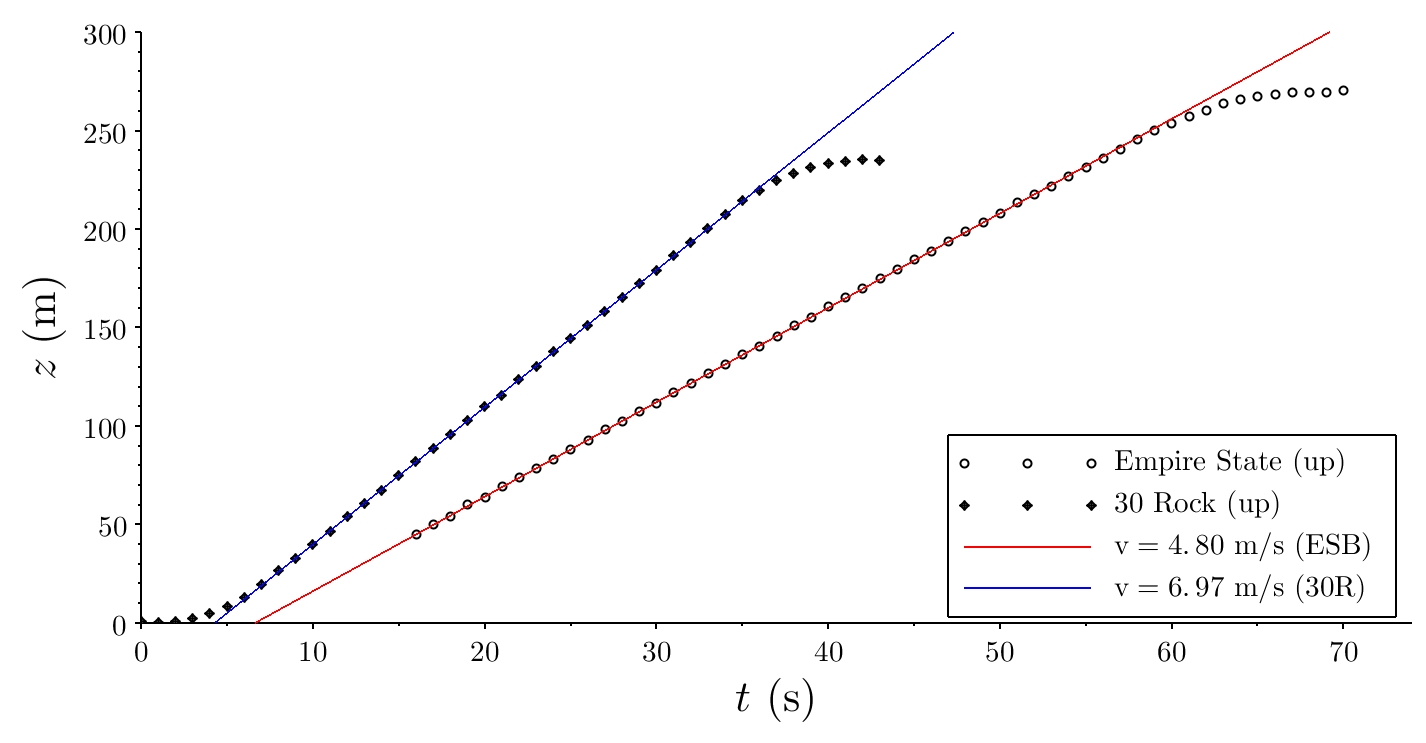}
\includegraphics[width=0.490\textwidth]{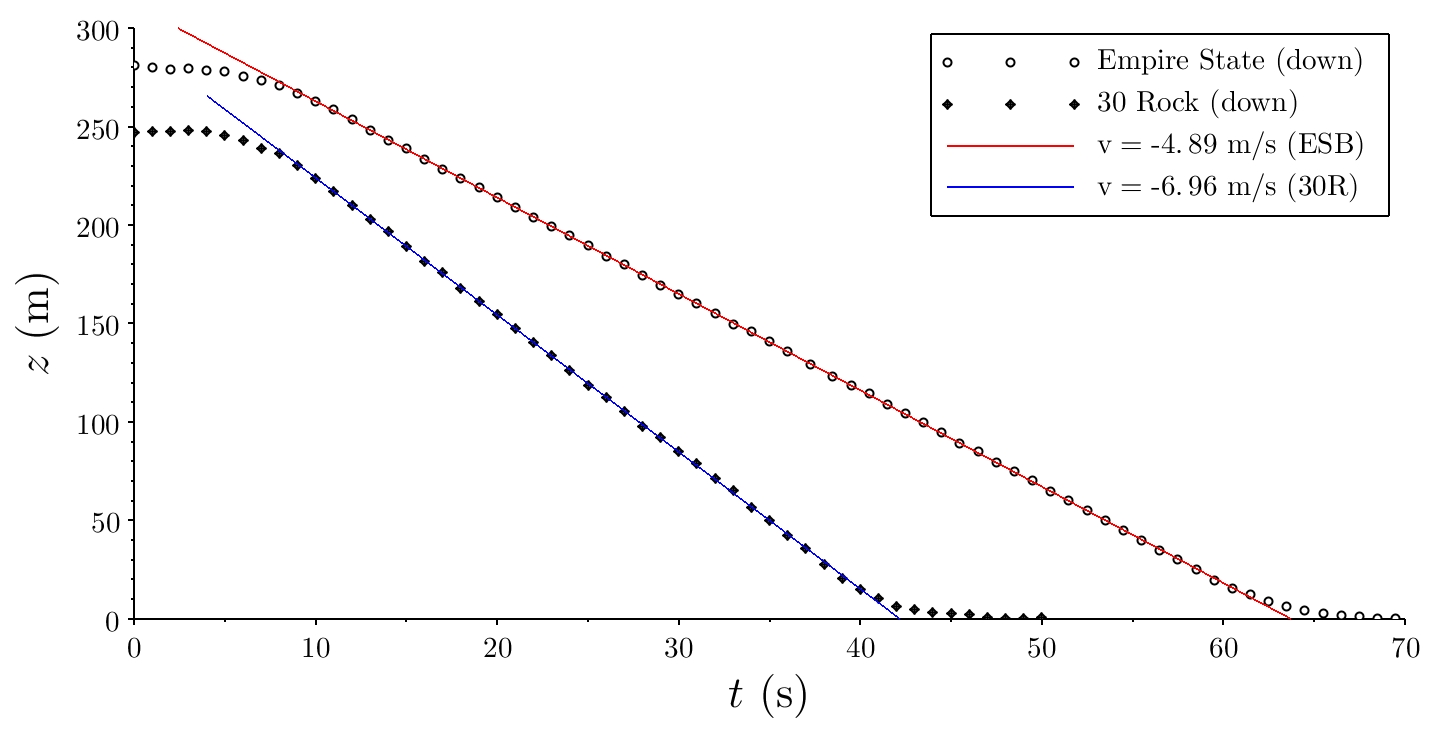}
\caption{Altitude of the elevators in the ESB and in 30R as a function of time, as obtained from the pressure values 
given in Fig.~\ref{fig:pre}.
Average vertical velocities are shown. }
\label{fig:alt}
\end{center}
\end{figure}

\subsection{Climbing stairways}

\begin{figure}
\begin{center}
\includegraphics[width=0.90\textwidth]{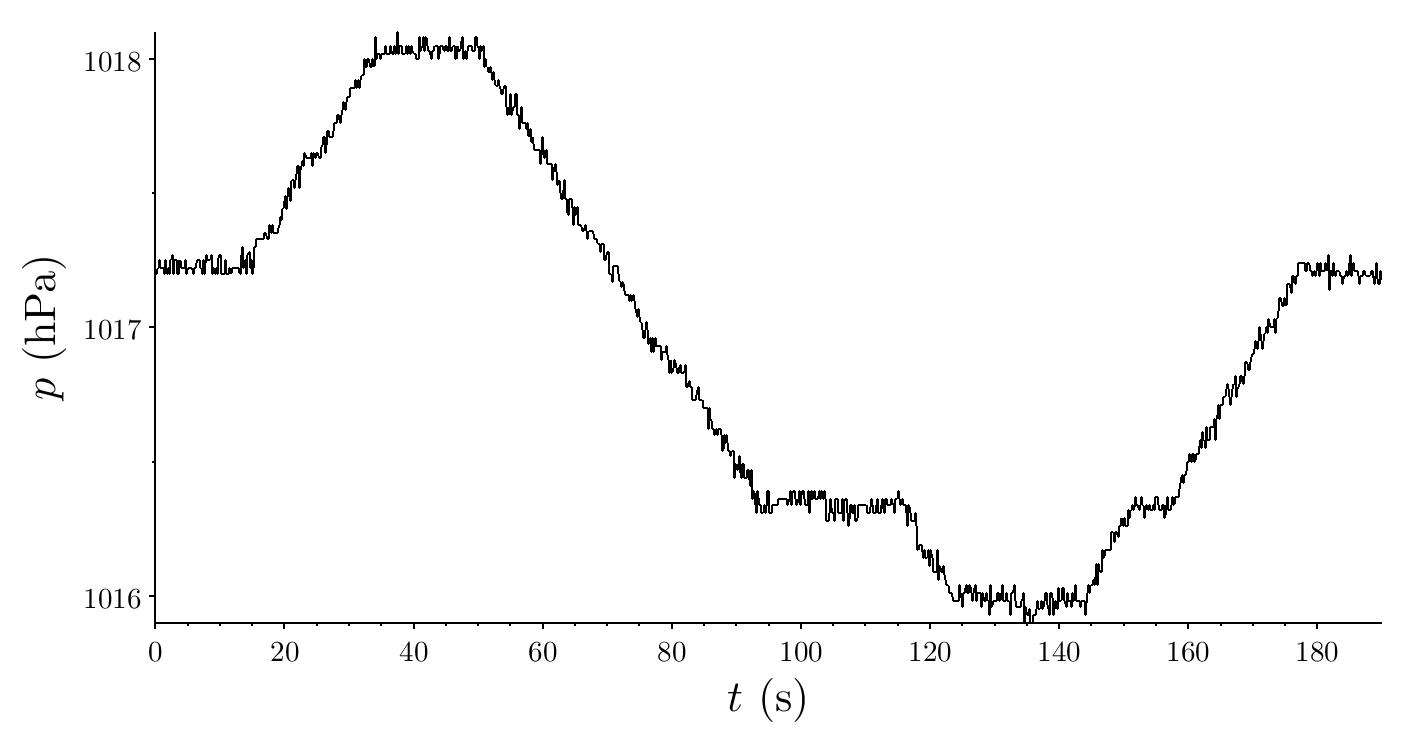}
\caption{Pressure values recorded by a smartphone held by 
a pedestrian in his/her hands during an excursion going across several stairways and corridors at the Universidad ORT Uruguay.
}
\label{fig:stairs}
\end{center}
\end{figure}

\begin{figure}
\begin{center}
\includegraphics[width=0.90\textwidth]{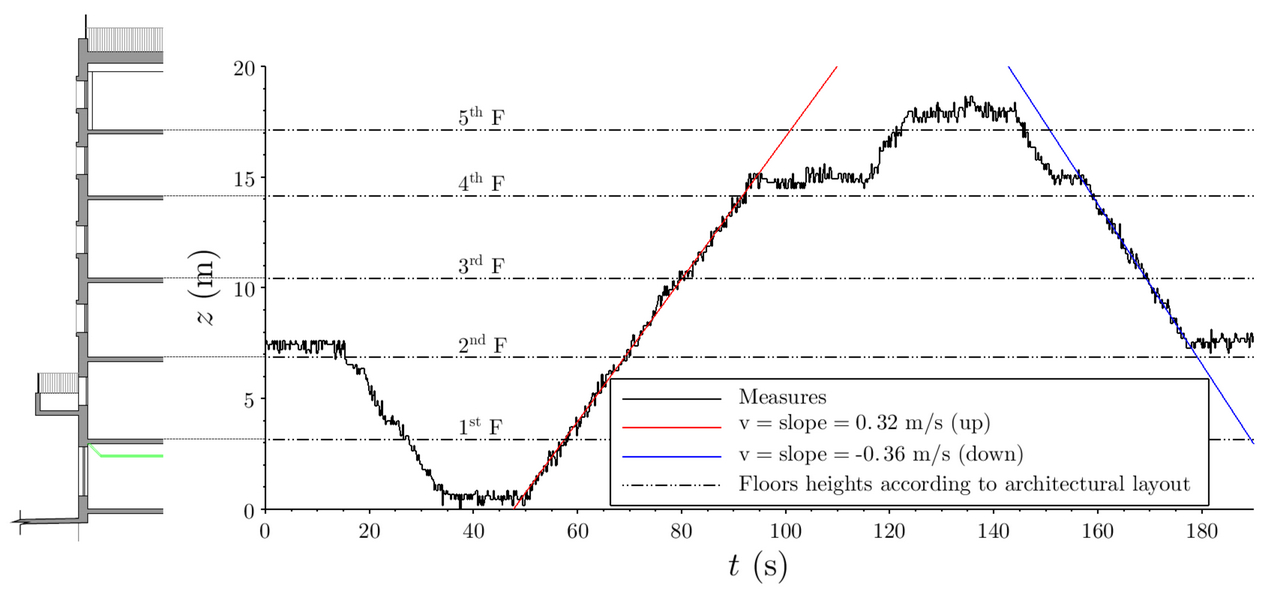}
\caption{Altitude of the pedestrian obtained from the pressure sensor, see Fig.~\ref{fig:stairs}, compared with the corresponding 
architectural layout during walking at the Universidad ORT Uruguay.
The experimenter started at the physics laboratory on the second floor, and descended to the ground floor, 
walked down a corridor,  took another stairway up to the fourth floor, walked down another corridor, 
and climbed stairs to the fifth floor. 
Then, she/he walked down another corridor and, finally, went down to the physics laboratory after a brief break at the fourth floor. 
The average vertical velocities in two sections, obtained using linear fits, are presented.
}
\label{fig:stairs-alt}
\end{center}
\end{figure}

The preceding results suggest that the previous experiments could be performed in an even simpler situation: 
a person climbing stairs while holding the smartphone. The following experiment was performed at
the Engineering School at the Universidad ORT Uruguay in Montevideo, starting and finishing at the Physics Laboratory after travelling through
several stairways and corridors. As expected, the pressure values measured during such excursion, shown in Fig.~\ref{fig:stairs},
are significantly more noisy than those registered  in  elevators (see Figs.~\ref{fig:fcien-pressure17} and  \ref{fig:pre}). However,  it is still possible to calculate the altitude,
as shown in Fig.~\ref{fig:stairs-alt}. In this situation, we must take into account the fact that the smartphone 
is held in the hands of the experimenter at
approximately $1$ m above the floor. After correcting for this height, a comparison of these results with the reference values derived from the architectural plans
indicated very good agreement.
Average values of the vertical velocity were estimated to be of the order of $1$ m at every $3$ s in two of the sections. 
This method for the estimation of velocity is also a good exercise for new students.

\subsection{Vertical velocity of a \textit{drone}}

\begin{figure}
\begin{center}
\includegraphics[width=0.550\textwidth]{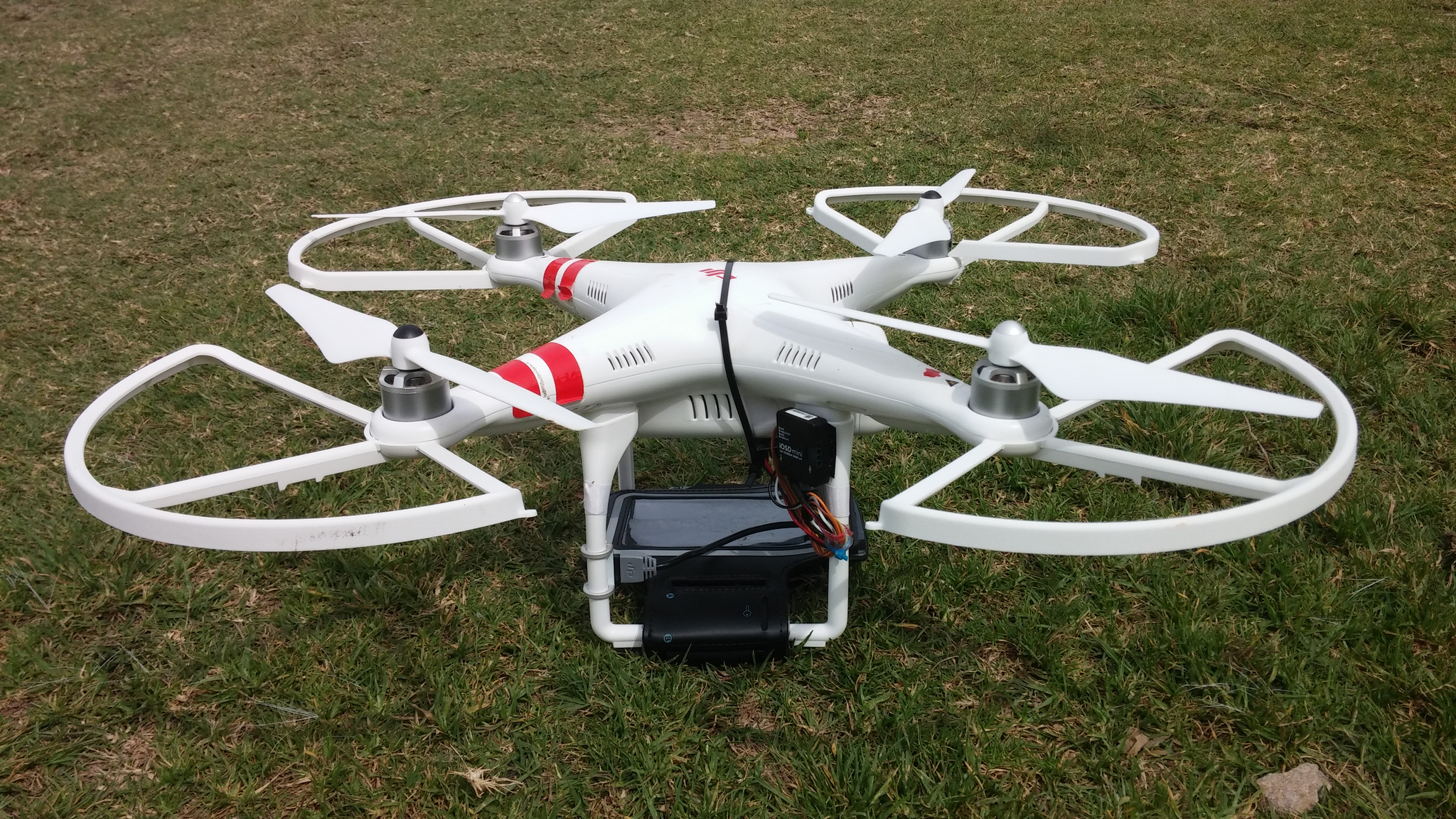}
\caption{Smartphone mounted on DJI Phantom 2 using an armband.}
\label{fig:drone}
\end{center}
\end{figure}

Remotely-controlled helicopters and planes have been used as toys for decades. However, recently, 
advances in sensor technologies have made it possible to easily and inexpensively fly and control these devices.
Along with their increasing availability, educational applications are also proliferating \cite{monteiro2015analyzing}. 
In this paper, we propose a simple experiment in which a smartphone is mounted on a DJI Phantom 2 using an armband case, as shown in Fig.~\ref{fig:drone}. 
 
\begin{figure}
\begin{center}
\includegraphics[width=0.950\textwidth]{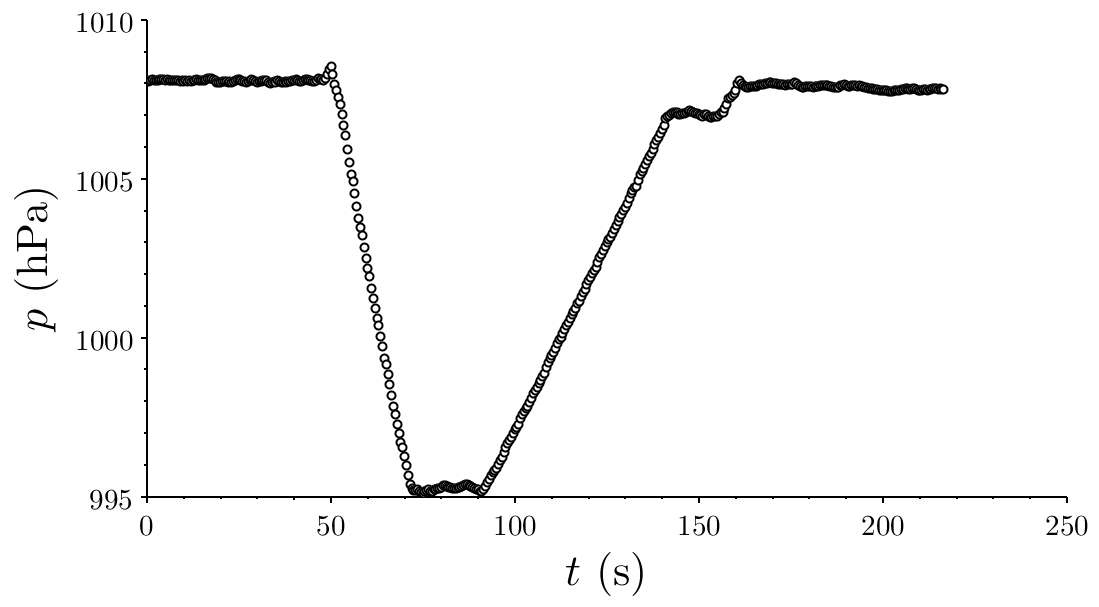}
\caption{Pressure as a function of time, as registered by a smartphone mounted on a drone during a simple flight. 
}
 \label{fig:drone-pres}
\end{center}
\end{figure}

The drone performed a simple flight while the smartphone was recording several variables using its built-in sensors, 
most importantly, the pressure sensor. First, it ascended to an altitude of  $120$ m relative to the take-off point (according to the telemetry of the drone's RF control),
remained hovering for approximately $20$ s and descended to an altitude  of $ 12$ m, and, finally, landed. 
During the flight, the AndroSensor app was used to record the atmospheric pressure and other variables of interest.
Pressure as a function of time is shown in Fig.~\ref{fig:drone-pres}.
According to the information provided by the manufacturer of the drone,
the maximum ascent speed is $6$ m/s (although, this value can vary a small amount based on the
attached weight) and the maximum descent speed is $2$ m/s.  In this experiment, care was taken to avoid harming people, animals, or property and to fulfill
all air traffic regulations. 

\begin{figure}
\begin{center}
\includegraphics[width=0.950\textwidth]{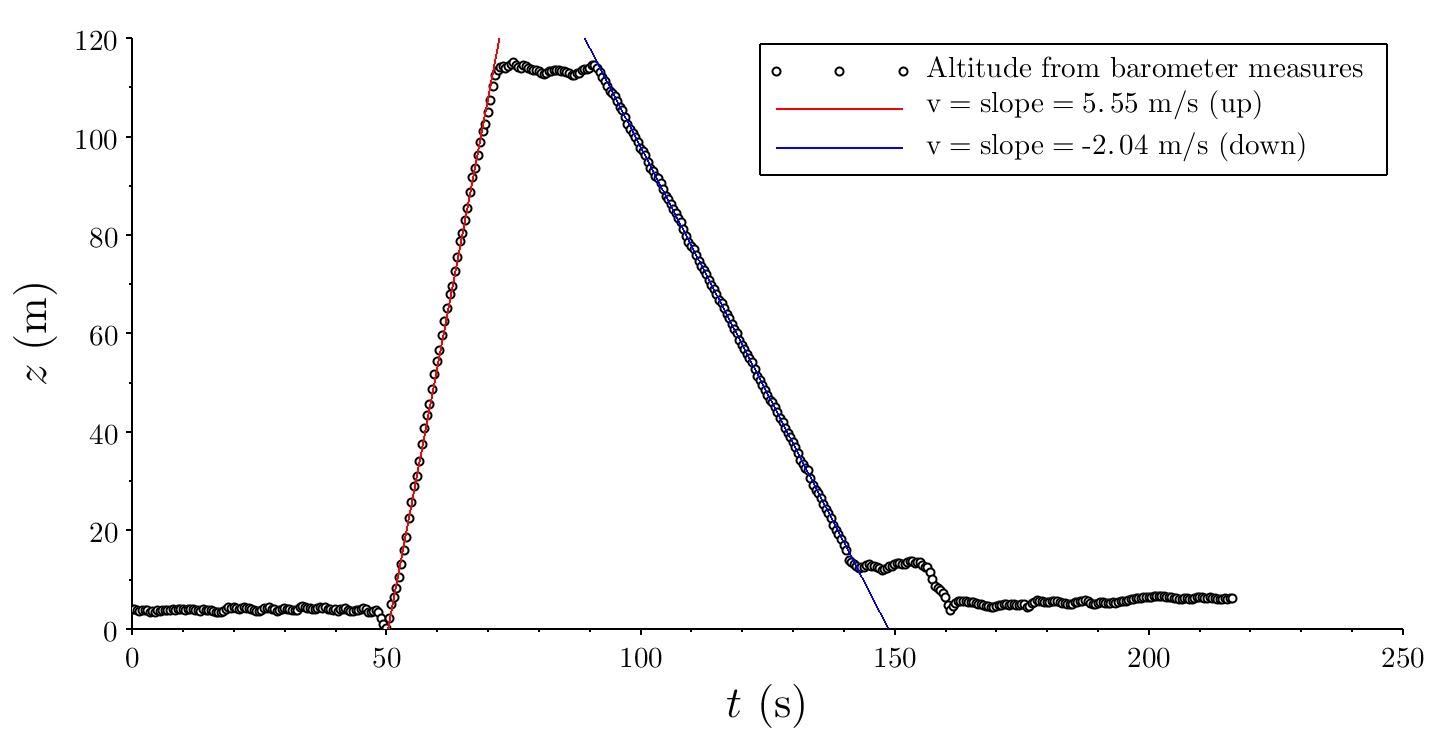}
\caption{Altitude of the drone as a function of time. The drone ascends to an altitude of about $115$ m, remained hovering for 20 s,
and descended to $10$ m before landing.
Ascending and descending vertical speeds are obtained using a linear fit.}
\label{fig:04}
\end{center}
\end{figure}

Figure \ref{fig:04} shows the altitude, as obtained using the pressure sensor and Eq.~\ref{eq:zP}. 
The ascending and descending speeds are obtained via linear
fits of the altitude during their respective time intervals. These values correspond well to the values reported by the manufacturer.
The acceleration measurements (not shown here) are significantly more noisy \cite{monteiro2015analyzing}.  As a consequence of the intrinsic noise of the device, numerical integration of the acceleration values does not provide good results for the calculations of velocity or position. 

\section{Concluding remarks and perspectives}
\label{sec:con}

In this paper, the use of smartphone pressure sensors was proposed to determine vertical velocities of elevators, pedestrians
climbing stairways, and drones. Using the hydrostatic approximation, which is valid in the first several hundred meters of the atmosphere,  
pressure is related to the altitude of the device and the vertical component of the velocity. 

We also measured the acceleration and obtained the velocity and altitude via numerical integration.
The results obtained using the pressure sensor are compared with those obtained using the smartphone's accelerometer. We showed
that data obtained using the pressure sensor are considerably less noisy. Because of the accumulation of error in the numerical integration,
the use of accelerometers is found to be less suitable than the use of the pressure sensor in this context.
The previous results were compared with measurements taken from the architectural plans of the buildings or information provided by the
manufacturer of the drone. Note that another popular sensor, GPS, which is present in most smartphones,
is not well-suited for this activity because it does not work well inside a building and the refresh rate is significantly lower.
In all cases, the comparisons validated the results obtained using the pressure sensor.  
One additional advantage is that this proposal is ideal as an external or outreach activity to allow students to gain insight into fundamental questions in fluid mechanics and thermodynamics.

\section{Acknowledgements}
We are grateful to Cecilia Cabeza and Cecilia Stari for enlightening discussions and to Ilan Cohn for his critical reading of the revised version.
We acknowledge PEDECIBA, Uruguay, for financial support.


\section*{References}


\begin{thebibliography}{10}

\bibitem{Chevrier2013}
Joel Chevrier, Laya Madani, Simon Ledenmat, and Ahmad Bsiesy.
\newblock Teaching classical mechanics using smartphones.
\newblock {\em The Physics Teacher}, 51:376, 2013.

\bibitem{monteiro2014angular}
Mart{\'\i}n Monteiro, Cecilia Cabeza, Arturo~C Marti, Patrik Vogt, and Jochen
  Kuhn.
\newblock Angular velocity and centripetal acceleration relationship.
\newblock {\em The Physics Teacher}, 52(5):312--313, 2014.

\bibitem{monteiro2014exploring}
Mart{\'\i}n Monteiro, Cecilia Cabeza, and Arturo~C Mart{\'\i}.
\newblock Exploring phase space using smartphone acceleration and rotation
  sensors simultaneously.
\newblock {\em European Journal of Physics}, 35(4):045013, 2014.

\bibitem{MONTEIRO2015}
Mart\'{\i}n Monteiro, Cecilia Cabeza, and Arturo~C. Marti.
\newblock {Acceleration measurements using smartphone sensors: Dealing with the
  equivalence principle}.
\newblock {\em {Revista Brasileira de Ensino de F\'i\-sica}}, 37:1303 --, 03
  2015.

\bibitem{countryman2014familiarizing}
Colleen~Lanz Countryman.
\newblock Familiarizing students with the basics of a smartphone's internal
  sensors.
\newblock {\em The Physics Teacher}, 52(9):557--559, 2014.

\bibitem{vieyrafive}
Rebecca Vieyra, Chrystian Vieyra, Philippe Jeanjacquot, Arturo Marti, and
  Mart{\'\i}n Monteiro.
\newblock Five challenges that use mobile devices to collect and analyze data
  in physics.
\newblock {\em The Science Teacher}, 82(9):32--40, 2015.

\bibitem{gonzalez2015doing}
Miguel~{\'A}ngel Gonz{\'a}lez~Rebollo, F{\'e}lix Huete, Juarez Bento~da Silva,
  {\'O}scar Mart{\'\i}nez, Diego Esteban, Willian Rochadel, Juan~Carlos
  Ca{\~n}edo, Javier Manso, and Manuel~{\'A} Gonz{\'a}lez.
\newblock Doing physics experiments and learning with smartphones.
\newblock 2015.

\bibitem{castro2013using}
Juan~Carlos Castro-Palacio, Luisberis Vel{\'a}zquez-Abad, Marcos~H Gim{\'e}nez,
  and Juan~A Monsoriu.
\newblock Using a mobile phone acceleration sensor in physics experiments on
  free and damped harmonic oscillations.
\newblock {\em American Journal of Physics}, 81:472, 2013.

\bibitem{Kuhn2013acoustic}
Jochen Kuhn and Patrik Vogt.
\newblock Analyzing acoustic phenomena with a smartphone microphone.
\newblock {\em The Physics Teacher}, 51:118, 2013.

\bibitem{monteiro2015measuring}
Mart{\'\i}n Monteiro, Arturo~C Marti, Patrik Vogt, Lutz Kasper, and Dominik
  Quarthal.
\newblock Measuring the acoustic response of helmholtz resonators.
\newblock {\em The Physics Teacher}, 53(4):247--249, 2015.

\bibitem{Forinash2012}
Kyle Forinash and Raymond~F. Wisman.
\newblock Smartphones as portable oscilloscopes for physics labs.
\newblock {\em The Physics Teacher}, 50:242, 2012.

\bibitem{0143-0807-36-6-065002}
Enrique Arribas, Isabel Escobar, Carmen~P Suarez, Alberto Najera, and Augusto
  Beléndez.
\newblock Measurement of the magnetic field of small magnets with a smartphone:
  a very economical laboratory practice for introductory physics courses.
\newblock {\em European Journal of Physics}, 36(6):065002, 2015.

\bibitem{lara2015demonstrations}
V O M Lara, D F Amaral, D~Faria, and L P Vieira.
\newblock Demonstrations of magnetic phenomena: measuring the air permeability
  using tablets.
\newblock {\em Physics Education}, 49(6):658, 2014.

\bibitem{kuhn2014iradioactivity}
Jochen Kuhn, Alexander Molz, Sebastian Gr{\"o}ber, and Jan Fr{\"u}bis.
\newblock iradioactivity—possibilities and limitations for using smartphones
  and tablet pcs as radioactive counters.
\newblock {\em The Physics Teacher}, 52(6):351--356, 2014.

\bibitem{moll2010amusement}
Rachel~F Moll.
\newblock An amusement park physics competition.
\newblock {\em Physics Education}, 45(4):362, 2010.

\bibitem{bagge2002classical}
Sara Bagge and Ann-Marie Pendrill.
\newblock Classical physics experiments in the amusement park.
\newblock {\em Physics Education}, 37(6):507, 2002.

\bibitem{vieyra2014analyzing}
Rebecca~E Vieyra and Chrystian Vieyra.
\newblock Analyzing forces on amusement park rides with mobile devices.
\newblock {\em The Physics Teacher}, 52(3):149--151, 2014.

\bibitem{cabeza2014learning}
Cecilia Cabeza, Nicolas Rubido, and Arturo~C Mart{\'\i}.
\newblock Learning physics at the water park.
\newblock {\em Physics Education}, in press(49):187--194, 2014.

\bibitem{kinser2015relating}
Jason~M Kinser.
\newblock Relating time-dependent acceleration and height using an elevator.
\newblock {\em The Physics Teacher}, 53(4):220--221, 2015.

\bibitem{monteiro2016exploring}
Martín Monteiro, Patrik Vogt, Cecilia Stari, Cecilia Cabeza, and Arturo~C.
  Marti.
\newblock Exploring the atmosphere using smartphones.
\newblock {\em The Physics Teacher}, 54(5):308--309, 2016.

\bibitem{vanini2016using}
Salvatore Vanini, Francesca Faraci, Alan Ferrari, and Silvia Giordano.
\newblock Using barometric pressure data to recognize vertical displacement
  activities on smartphones.
\newblock {\em Computer Communications}, 87:37--48, 2016.

\bibitem{carvalho2014observing}
Luiz Raimundo Moreira~de Carvalho and Helio Salim~de Amorim.
\newblock Observing the atmospheric tides: an application of the arduino board
  with sensors for barometric pressure and temperature.
\newblock {\em Revista Brasileira de Ensino de F{\'\i}sica}, 36(3):1--7, 2014.

\bibitem{silva2015open}
RB~Silva, LS~Leal, LS~Alves, RV~Brand{\~a}o, RCM Alves, EV~Klering, and
  RP~Pezzi.
\newblock Open source weather stations: A research and technological
  development project.
\newblock {\em Revista Brasileira de Ensino de F{\'\i}sica}, 37(1):1505, 2015.

\bibitem{muralidharan2014barometric}
Kartik Muralidharan, Azeem~Javed Khan, Archan Misra, Rajesh~Krishna Balan, and
  Sharad Agarwal.
\newblock Barometric phone sensors: More hype than hope!
\newblock In {\em Proceedings of the 15th Workshop on Mobile Computing Systems
  and Applications}, page~12. ACM, 2014.

\bibitem{kuhn2014analyzing}
Jochen Kuhn, Patrik Vogt, and Andreas M{\"u}ller.
\newblock Analyzing elevator oscillation with the smartphone acceleration
  sensors.
\newblock {\em The Physics Teacher}, 52(1):55--56, 2014.

\bibitem{monteiro2015analyzing}
Mart{\'\i}n Monteiro, Cecilia Stari, Cecilia Cabeza, and Arturo~C Marti.
\newblock Analyzing the flight of a quadcopter using a smartphone.
\newblock {\em arXiv preprint arXiv:1511.05916}, 2015.

\end{thebibliography}
\end{document}